\documentclass[twocolumn,nofootinbib,amsmath,amssymb,aps,prd,balancelastpage,superscriptaddress]{revtex4-1}

\usepackage{color}
\usepackage[dvipsnames]{xcolor}
\usepackage[active]{srcltx}
\usepackage{amsmath,amsfonts,amssymb,amsthm,amstext,amscd,eucal,srcltx}
\usepackage{epsfig,graphicx,bm}
\usepackage{epstopdf, epsf}
\usepackage{dcolumn}
\usepackage{hyperref}
\usepackage{tensor}
\usepackage{lipsum}
\usepackage{appendix}
\usepackage{tikz}
\usepackage{caption}
\usepackage{subcaption}

\newcommand{\be}{\begin{equation}}
\newcommand{\ee}{\end{equation}}

\newcommand{\bse}{\begin{subequations}}
\newcommand{\ese}{\end{subequations}}
\newcommand{\bea}{\begin{eqnarray}}
\newcommand{\eea}{\end{eqnarray}}
\newcommand{\ba}{\begin{array}}
\newcommand{\ea}{\end{array}}
\newcommand{\bc}{\begin{center}}
\newcommand{\ec}{\end{center}}

\begin{document}
\preprint{IPM/P-2012/009}
\vspace*{3mm}

\title{QCD surprises: strong CP problem, neutrino mass, Dark Matter and Dark Energy}

\author{Andrea Addazi}
\email{addazi@scu.edu.cn}
\affiliation{Center for Theoretical Physics, College of Physics Science and Technology, Sichuan University, 610065 Chengdu, China}
\affiliation{INFN sezione Roma {\it Tor Vergata}, I-00133 Rome, Italy, EU}

\author{Antonino Marcian\`o}
\email{marciano@fudan.edu.cn}
\affiliation{Center for Field Theory and Particle Physics \& Department of Physics, Fudan University, 200433 Shanghai, China}
\affiliation{Laboratori Nazionali di Frascati INFN, Frascati (Rome), Italy, EU}

\author{Roman Pasechnik}
\email{Roman.Pasechnik@thep.lu.se}
\affiliation{Department of Astronomy and Theoretical Physics, Lund University, Solvegatan 14A S 223 62 Lund, Sweden, EU}

\author{Kaiqiang Alan Zeng}
\affiliation{Center for Theoretical Physics, College of Physics Science and Technology, Sichuan University, 610065 Chengdu, China}

\begin{abstract}
\noindent
An unexpected explanation for neutrino mass, Dark Matter (DM) and Dark Energy (DE) from genuine Quantum Chromodynamics (QCD) of 
the Standard Model (SM) is proposed here, while the strong CP problem is resolved without any need to account for fundamental axions. 
We suggest that the neutrino sector can be in a double phase in the Universe: i) relativistic neutrinos, belonging to the SM; ii) non-relativistic 
condensate of Majorana neutrinos. The condensate of neutrinos can provide an attractive alternative candidate for the DM, being in a cold 
coherent state. We will explain how neutrinos, combining into Cooper pairs, can form collective low-energy degrees of freedom, hence 
providing a strongly motivated candidate for the QCD (composite) axion. 
\end{abstract}

\maketitle


{\bf Introduction}.
The idea that Dark Energy (DE) in the form of cosmological constant (CC) can originate from the Quantum Chromodynamics (QCD) hadron sector 
of the Standard Model (SM) was first proposed by {\it Zeldovich} \cite{Zeldovich:1967gd} and {\it Sakharov} \cite{Sakharov:1967pk}, and further elaborated 
in Ref.~\cite{Pasechnik:2013poa}. This is based on an interesting coincidence that the following combination of QCD and Planck scales 
\begin{equation}
\label{epsilonLambda}
\epsilon_{\Lambda}\simeq \frac{1}{(2\pi)^{4}}\frac{m_{\pi}^{6}}{M_{Pl}^{2}}\simeq  3\times 10^{-35}\, {\rm MeV}^{-4}\, , 
\end{equation}
appears to be remarkably close to the measured value of the CC~\cite{Hinshaw:2012aka,Ade:2015rim,Ade:2015xua}
\begin{equation}
\label{expe}
\epsilon_{\Lambda, {\rm exp}}=3.0\pm 0.7\, \times 10^{-35}\,  {\rm MeV}^{-4}\, .
\end{equation}
Above, $\epsilon_{\Lambda}$ is the vacuum energy density, $m_{\pi}$ is the pion mass and $M_{Pl}$ is the Planck scale. In Ref.~\cite{Pasechnik:2013poa} it was 
explicitly demonstrated that such $\epsilon_{\Lambda}$ term naturally emerges as a non-vanishing first-order gravitational correction to the QCD vacuum 
energy density. The crucial condition is that the average quark and gluon condensates' energy density vanishes in the infrared limit i.e. when the averaging 
over comoving volume is performed beyond the Fermi scale of QCD. 

Another intriguing coincidence is that the scale of cosmological DE is close to ${\rm meV}^{4}$, which resembles the (forth power of the) lightest neutrino mass scale. As was suggested by {\it Ginzburg} and {\it Zharkov} back in 1967 \cite{GZ}, it is possible that neutrinos can be non-relativistically produced in a superfluid phase, related to the neutrino mass gap and DE. Such a proposal was elaborated later on in Refs.~\cite{Alexander:2009uu,Alexander:2009yb,
Alexander:2013baa,Dvali:2016uhn,Addazi:2016oob} concluding that neutrino may play a crucial role in the formation of the observable DE.

On the other hand, the fact that the Dark Matter (DM) and Baryon Matter (BM) in the Universe have relatively close abundances, i.e. $\Omega_{\rm DM}/\Omega_{\rm BM}\simeq 5$, 
might hint towards their common origin related to QCD dynamics, especially, having in mind that most of the proton mass ($95\%$ or so) is sourced by QCD gluon 
binding energy. One may conclude that if DM has a completely different origin and has been created at a totally different energy scale than that 
for the BM, such a remarkable order-of-magnitude coincidence of their abundances would seem largely unnatural and rather accidental.

Are the above coincidences the matter of a numerical accident, or is there any possible unification principle behind these? 

\vspace{0.1cm}

We show that a tiny mass scale of the SM active neutrino species, as well as the DM and the DE components of the Universe, can be traced back to a common unified origin related to the strong (QCD) sector 
of the SM\footnote{In Ref.~\cite{Addazi:2016oob}, an alternative approach has been considered where a new non-linear dark photon interaction unifying neutrino mass scale and DE was postulated.}.The neutrino condensation is catalysed by its interaction with the DE, which is in turn originated from the residual QCD sector arising from an emergent mirror symmetry (MS) in the QCD vacuum broken by gravitational effects \cite{Pasechnik:2013poa}. The CC is protected from receiving large contributions from the QCD condensates by means of the MS --- see e.g.~Refs.~\cite{Addazi:2019mlo,Addazi:2018ctp} for more discussions on this phenomenon. In our picture, we postulate that neutrinos have a Peccei-Quinn (PQ) scale suppressed portal with the strongly-coupled QCD sector through quantum anomalies. The QCD axion is then identified with the composite Cooper pair of neutrinos\footnote{In companion papers, we theorised the existence of a dark QCD-like sector, so as to unveil DE \cite{Addazi:2016nok,Addazi:2018xoa,Addazi:2020tbk} and eliminate the QCD vacuum contribution to the DE \cite{Pasechnik:2013sga,Pasechnik:2016twe}. Nonetheless, here we go beyond this assumption and explore the full potential of non-perturbative QCD to address the missing matter/energy problem in the Universe.}. On the other hand, the neutrino condensate can naturally account for the right amount of DM (for a recent discussion of the sterile neutrino condensate and possible connections to the DM, see Ref.~\cite{Xue:2020cnw}). Indeed, the issue is automatically solved if the lowest excitation of the neutrino condensate is identified with the QCD axion, being produced from the misalignment mechanism in the early Universe in a non-relativistic cold phase. More specifically, the neutrino condensation dynamically breaks the $U_{\rm PQ}(1)$ symmetry\footnote{Regarding the vacuum stability, a solution to the problem was proposed 
by e.g.~invoking a new Naturalness principle, the $\mathcal{H}$olographic $\mathcal{N}$aturalness \cite{Addazi:2020hrs,Addazi:2020wnc,Addazi:2020mnm}, 
and thus this aspect is not further discussed here. On the other hand, the screening of the QCD vacuum energy density outside the Fermi scale of QCD 
was explored in Ref.~\cite{Addazi:2019mlo}, having possible phenomenological implications in gravitational waves astrophysics \cite{Addazi:2018ctp}. 
}. \\

{\bf The model}. 
We may start considering a complex scalar field that is charged with respect to the PQ symmetry, and is emerging from a BCS-like condensation of neutrinos, namely,
\begin{equation}
\label{Phu}
\Phi=\rho \, e^{\imath \frac{a}{f_{\rm PQ}}}\,,\qquad a=\frac{1}{f_{a}^2}\nu^{T}{\mathcal C}^{-1}\gamma_{5}\nu\,,
\end{equation}
where $f_{\rm PQ}$ is the PQ scale, $f_{a}$ is the axial-symmetry breaking scale (the axion decay constant) related to the neutrino condensation scale --- this is in turn related, 
as we will see later on, to the QCD interaction portal. The $a$-field is introduced as an auxiliary field able to embed a Majorana-like neutrino operator. 


The PQ breaking Lagrangian is expressed as
\begin{equation}
\label{LPp}
\mathcal{L}=\partial \Phi^{\dagger}\partial \Phi - V(\Phi)\, , \quad V(\Phi)=(|\Phi|^{2}-f_{\rm PQ}^{2})^{2}\,,
\end{equation}
where $V(\Phi)$ acquires a VEV at the $f_{\rm PQ}$ scale. 
The neutrino current can be considered as the imaginary part of a scalar field, interacting with 
its real component. Accounting for the Sombrero potential, the mixing terms cast
\begin{equation}
\label{Lmix}
\mathcal{L}_{\rm mix}=\frac{1}{f_{\rm PQ}^2}\rho^{2}(\partial a)^{2}\, . 
\end{equation}
In terms of the neutrino fields, the operator in \eqref{Lmix} can be straightforwardly recast using the identity
\begin{align*}
(\partial_{\mu} a)^{2} =\, \, &
 \frac{1}{f_{a}^{4}}(\partial \nu^{T} \mathcal{C}^{-1})^{2}\nu^2+\frac{1}{f_{a}^{4}}  (\nu^{T}\mathcal{C}^{-1})^{2}(\partial \nu)^{2}\\
&+2\frac{1}{f_{a}^{4}}(\partial_{\mu} \nu^{T}\mathcal{C}^{-1}\nu)( \nu^{T}\mathcal{C}^{-1}\partial^{\mu}\nu)\, .
\end{align*}

\begin{figure}[ht]
\centerline{ \includegraphics [width=1.0\columnwidth]{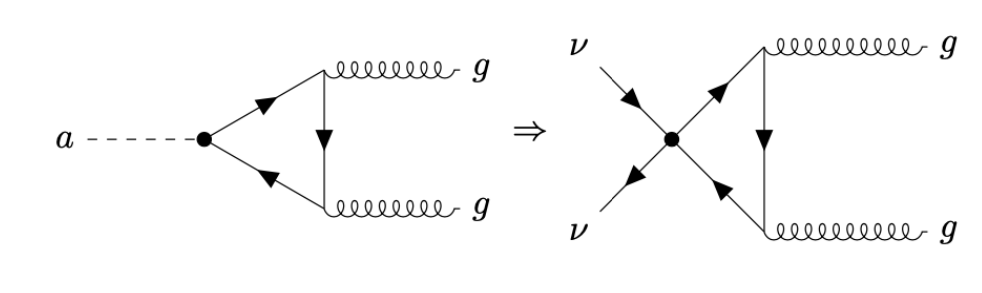}}
\caption{The QCD axion portal with the gluon field through the axial anomaly is replaced by a four-fermion interaction between the SM neutrinos and the quarks.}
\label{fig:pep}
\end{figure}

The spontaneous symmetry breaking of the PQ symmetry cannot generate any mass for the SM neutrino pairs. Nonetheless, the neutrino can acquire a mass gap by means of the axial anomaly induced by the QCD sector\footnote{In order to compute the axial anomaly effect, the axion coupling to the internal fermion line is replaced by an effective four-fermion coupling between SM neutrinos and quark lines in the triangle loop. Such an effective operator can be UV-completed considering e.g.~beyond-the-SM scalar or vector lepto-quarks that can mediate tree-level interactions between neutrinos and quarks at high energy scales. In the SM framework, the $Z$-boson exchange can also mediate such four-fermion interactions at tree level. However, this does not provide a mixing of different neutrino flavours that may be induced at a loop level only. Possible UV completion for such an effective operator is not relevant for the current considerations and can be a subject for a future dedicated work.}. This is possible if the neutrino has an effective coupling to the field strength of the type
\be
\frac{1}{f_{\rm PQ}f_{a}^{2}}(\nu^{T}{\mathcal C}^{-1}\gamma_{5}\nu)\, G\widetilde{G}\, ,
\label{eq:1}
\ee
which is effectively corresponding to the coupling of the composite axion,
$f_{\rm PQ}^{-1}a\, G\widetilde{G}$ (see Fig.~\ref{fig:pep}).
\begin{figure}[ht]
\centerline{ \includegraphics [width=1.0 \columnwidth]{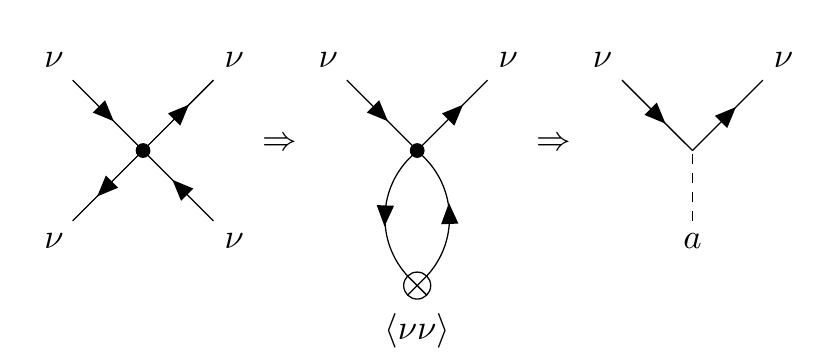}}
\caption{The four-fermion coupling is responsible for condensation of a neutrino pair, spontaneously breaking the chiral symmetry and generating the neutrino mass scale in the SM. The neutrino pair in such a condensate can then be identified with a QCD axion effectively coupled to the neutrinos.}
\label{fig:mass}  
\end{figure}

Thanks to this coupling, neutrinos can be produced in pairs, as cold, during the QCD phase transition. They can condense as 
non-relativistic Cooper pairs\footnote{The SU(2) group-coherent nature of BCS-like condensates, and their phenomenological 
implications also for early cosmology, have been investigated in Ref.~\cite{Dona:2016fip}.}, and be produced, as mentioned above, 
from the misalignment mechanism --- the very same one as for the fundamental axions \cite{Zyla:2020zbs}. (For a shorter notation, 
we will denote $\nu^{T}\mathcal{C}^{-1}\gamma_{5}\nu$ with $\nu\nu$ in what follows.) The Cooper pair mass has a functional form 
\be
m_{a}^{2}=\Big\langle \frac{\partial^{2}V_{\rm eff}}{\partial a^{2}}\Big\rangle\Big|_{\langle a \rangle}\, , 
\label{eq:3}
\ee
where $a$ is an auxiliary variable representing the neutrino (pseudo-)scalar pair. 
As it is well known, the effective axion potential is generated by non-perturbative QCD effects
such as QCD instantons. Thus, the axion mass term is given by the vacuum expectation value 
of the topological QCD term as follows
\be
m_{a}^{2}=\frac{1}{32\pi^{2}f_{\rm PQ}}\frac{\partial}{\partial a}\langle G\widetilde{G}\rangle \Big|_{\langle a\rangle}\, . 
\label{eq:4}
\ee
The gluon-condensate quantum oscillation may trigger creation of neutrinos' pairs, hence inducing the higher-order 
operator which provides an effective coupling with $a$. Since the axion field emerges as a condensate of cold neutrinos, 
the mass of the axion shall be evaluated at the minimum of the effective potential in the field configuration space, namely,
\be
\Big \langle \frac{\partial V_{\rm eff}}{\partial a}\Big \rangle \Big|_{\langle a \rangle}=\frac{1}{32\pi^{2}}\frac{1}{f_{\rm PQ}}
\langle G\widetilde{G}\rangle\Big|_{\langle a \rangle}=0\, . 
\label{eq:5}
\ee
As it happens for the standard axion, the $a(\nu)$ state has a periodic shift symmetry, 
with periodicity 
$$\theta-\frac{1}{f_{\rm PQ}f_{a}^2}\langle \nu\nu\rangle\, ,$$
forcing the neutrino pair to acquire an expectation value
$$\langle \nu\nu\rangle=f_{a}^{2}f_{\rm PQ}\theta\, .$$

Within the instanton gas approximation, the effective potential acquires the form 
\be
V_{\rm eff}(\nu)\simeq K\cos\Big(\frac{\nu\nu}{f_{\rm PQ}f_{a}^{2}}\Big)\, , 
\label{eq:6}
\ee
$K$ being of order $\Lambda_{\rm QCD}^{4}$. The composite axion mass can be related to the quark mass matrix 
and the QCD scale by means of 
\be
m_{a}^{2}=\frac{\mathcal{A^2}}{f_{\rm PQ}^{2}}\frac{\mathcal{V}K}{\mathcal{V}+K\, {\rm Tr}\, \mathcal{M}^{-1}}\, , 
\label{eq:7}
\ee
where $\mathcal{A}$ is the color anomaly of the $U_{\rm PQ}(1)$ current ($\mathcal{A}=1$ for $N_{f}=N_{c}$), $K$ is related to the effective potential $V_{\rm eff}$ (as specified above), $\mathcal{M}$ is the quark mass matrix, and $\mathcal{V} \sim \langle \bar{q}q\rangle\sim \Lambda_{\rm QCD}^{3}$. 

The neutrino pair acquires a small mass mixing with the QCD pseudo-Goldstone bosons $\pi,\eta$. Neglecting 
the contribution of the $s$-quark in the first approximation, the neutrino-pair mass formula recasts  
\be
m_{a}=\frac{\mathcal{A}\, \zeta^{1/2}}{1+\zeta}\frac{f_{\pi}}{f_{a}}m_{\pi} \, \simeq \, 
\mathcal{A}\,\Big(\frac{10^{9}\, {\rm GeV}}{f_{a}}\Big){\rm meV}\, ,
\ee
with 
\be 
(m_{u}+m_{d})\langle \bar{q}q\rangle=m_{\pi}^{2}f_{\pi}^{2}\, ,
\ee
$\zeta=m_{u}/m_{d}\simeq 0.6$ and $m_{\pi}$, $f_{\pi}$ being, respectively, the pion mass and the pion decay constant. 
Furthermore, the resulting mass gap of the neutrino pair can be related to the composite axion mass originating 
from the QCD axial anomaly.

A Nambu-Jona-Lasinio mechanism for neutrinos' condensation can be envisaged, accounting for the generation of a mass gap 
triggered by an effective four-fermion interaction term of the form
\be
G_{\nu}\, \nu\nu\nu\nu\,  \rightarrow  G_{\nu} \langle \nu \nu\rangle  \nu\nu \,,
\label{eq:8}
\ee
where 
\be G_{\nu}|\langle \nu\nu\rangle|=m_{a}\, , \qquad  |\langle \nu\nu\rangle|=f_{a}^{2}f_{\rm PQ}\theta\, .
\ee
Hence, the effective coupling $G_{\nu}$ here is highly suppressed as
\be
G_{\nu}=m_{a}/f_{a}^{2}f_{\rm PQ}\theta  \,,
\ee
which is several orders of magnitude higher than cosmological limits on four-interaction neutrino terms
(see e.g. Ref.~\cite{Cyr-Racine:2013jua}).

Due to \eqref{eq:8}, for energies around the meV-scale, a neutrino-meson theory emerges accounting for several effective degrees of freedom. 
The axion is produced as an analogue of the $\eta'$-meson in QCD, namely, $a\equiv \eta_{\nu}'$. A small SM neutrino mass scale 
can be generated through the QCD conformal anomaly term starting from Fig.~\ref{fig:pep} and sinking the gluon legs into 
the QCD gluon condensate. This would provide a contribution to the neutrino mass related to the axial anomaly. 
Replicating the same mechanism for the three generations of neutrinos, we derive the form of the neutrino mass matrix 
\be
\mathcal{M}_{ij}\nu_{i}\nu_{j}=G_{\nu\, ij}\langle \nu_{i}\nu_{j}\rangle\, \nu_{i}\nu_{j}\, , 
\ee
$G_{\nu\,  ij}=(m_{a})_{ij}/f_{a}^{2}f_{\rm PQ}\theta$ now encoding the relevant couplings of the neutrinos to the QCD sector as 
\be
\frac{1}{f_{a}^{2}f_{\rm PQ}}y_{ij}\nu_{i}\nu_{j}G\tilde{G}\, . 
\ee

In the current approach, the composite axion mass is directly related to the neutrino mass scale. Therefore, if we aim at explaining both the neutrino oscillations and the cosmological bounds on the neutrino sector, we shall assume that the composite axion has necessarily a mass within the range $1\div 10^{3}\, {\rm meV}$. 
This is certainly a sharp prediction for the axion composite state of neutrinos, which imposes a certain rigid range for the PQ scale too while invoking the composite axion as an explanation for the cosmological DM. 

\vspace{0.2cm}

{\bf QCD and Dark Energy}. In Refs.~\cite{Addazi:2019mlo,Addazi:2018ctp}, the possibility of a non-perturbatively emergent $\mathbb{Z}_{2}$ MS between 
chromoelectric (CE) and chromomagnetic (CM) condensates in QCD was envisaged. Such a MS can explain why the QCD vacuum energy density averaged 
over the Fermi scale volume is below the measured CC value in the Universe. In other words, the QCD conformal anomaly gets to vanish outside the Fermi 
scale, while $\Lambda_{\rm QCD}$ makes only sense inside the Fermi world as an energetic characteristics of each of the two condensates. Now, we would 
expect that quantum gravity effects would break every global symmetries in Nature including the MS of QCD~\cite{Banks:2010zn,Vafa:2005ui,ArkaniHamed:2006dz}. 

As a review of the concepts behind the mirror QCD symmetry, one can start from the effective Lagrangian 
of a pure Yang-Mills theory (such as QCD gluodynamics) \cite{Pagels:1978dd}
\be
\mathcal{L}_{\rm eff}=\frac{\mathcal{J}}{4\bar{g}^{2}}\,,\,\,\,\,\,\,\, \bar{g}\equiv \bar{g}(\mathcal{J})\,,\,\,\,\,\,\,\,\, 
\mathcal{J}=-\frac{\mathcal{F}_{\mu\nu}^{a}\mathcal{F}^{\mu\nu}_{a}}{\sqrt{-g}}\,, 
\label{eq:21}
\ee
where $g\equiv {\rm det}(g_{\mu\nu})$ and for a FLRW geometry $g_{\mu\nu}=a^{2}(\eta){\rm diag}(1,-1,-1,-1)$. 
Eq.~\eqref{eq:21} is complemented by the renormalization flow equation 
\be
\frac{d{\rm ln}|\bar{g}|^{2}}{d{\rm ln}|\mathcal{J}|/\mu_{0}^{4}}=\frac{\beta}{2}\, , 
\label{eq:22}
\ee
where $\beta\equiv \beta(\bar{g}^{2})$ is the $\beta$-function while $\mu_{0}$ denotes the infrared cutoff scale. 
The considered effective action approach is based on the Savvidy vacuum model \cite{Savvidy:1977as,Matinyan:1976mp}, 
which is an effective method for describing the ground state dynamics in quantum Yang-Mills theories at long distances.
The equation of motion of the considered theory (such as gluodynamics of QCD) reads
\begin{eqnarray}
&&\mathcal{D}_{\nu}^{ab}\Big[\frac{\mathcal{F}_{b}^{\mu\nu}}{\bar{g}^{2}\sqrt{-g}}\Big(1-\frac{1}{2}\beta(\bar{g}^{2}) \Big) \Big]=0\, , \\
&&\mathcal{D}_{\nu}^{ab}\equiv \Big( \delta^{ab}\frac{\partial_{\nu}(\sqrt{-g})}{\sqrt{-g}}-f^{abc}\mathcal{A}^{c}_{\nu}\Big)\, .
\end{eqnarray}
The above equations entail the exact ground-state solution 
\be
\beta(\bar{g}_{*}^{2})=2, \qquad \mathcal{J}^{*}>0\, , 
\label{eq:26}
\ee
corresponding to a CE condensate. The $\mathbb{Z}_{2}$ mirror symmetry of the effective action concerns simultaneous 
permutations of the field strength and couplings as follows
\be
\mathcal{J}^{*}\rightarrow -\mathcal{J}^{*},\qquad \bar{g}^{2}_{*}\rightarrow -\bar{g}_{*}^{2},\qquad \beta\rightarrow -\beta\, . 
\label{eq:29}
\ee
As shown in Refs.~\cite{Addazi:2019mlo,Addazi:2018ctp}, the CE condensate 
has a $\mathbb{Z}_{2}$ mirror ``partner'' -- the CM condensate -- which is a solution of the equation 
\be
\mathcal{D}_{\nu}^{ab}\frac{\mathcal{F}_{b}^{\mu\nu}}{\bar{g}^{2}\sqrt{-g}}=0 \,.
\label{eq:27}
\ee
The CE and CM condensates provide equal-magnitude but opposite-sign energy density contributions to the ground state 
reached asymptotically in the infrared regime of the theory, i.e.
\be
\epsilon_{\rm vac}\equiv \frac{1}{4}\langle T_{\  \nu}^{\nu}\rangle_{\rm vac}=\mp \, \mathcal{L}_{\rm eff}(\mathcal{J}^{*})\,,
\label{eq:28}
\ee
hence restoring the conformal symmetry of the theory at large spacetime separations.

Quantum gravity symmetry-breaking effects may involve effective operators suppressed by powers of the Planck scale. 
At the same time, effective QCD operators in the presence of gravity trigger a soft breaking of the mirror symmetry 
at the $\Lambda_{\rm QCD}$ scale through gravitational interactions, hence spoiling 
the cancellation mechanism among CE and CM condensates. A relevant example illustrating such an effect can be given as follows
\begin{equation}
\label{ma}
\mathcal{L}_{\rm breaking}=\frac{1}{M_{Pl}^{4}}(FF)^{2}+\dots
+ \frac{1}{M_{Pl}^{2}}(\bar{q}q)^{2}+\dots\, ,
\end{equation}
where the last operator can be connected to the accumulation of a vacuum CE flux-tube energy in the leading order, i.e.
\begin{equation}
\label{epp}
\epsilon_{\Lambda}\propto\frac{\Lambda_{\rm QCD}^{6}}{M_{Pl}^{2}}\,. 
\end{equation}
A rigorous semi-classical calculation of the QCD contribution to the cosmological constant yielding 
the same scaling with the QCD and Planck scales can be found in Ref.~\cite{Pasechnik:2013poa}.
As was mentioned above, there is an intriguing coincidence between this value of the vacuum energy 
density founds as a gravity-induced infrared-residual of the QCD vacuum density and the measured value of the CC.\\
 
{\bf Conclusions}. We have shown a novel simple mechanism to simultaneously resolve the Dark Matter, Dark Energy, QCD-CP and neutrino mass problems, in one striking shot. The neutrinos can acquire a tiny mass gap through a QCD anomalous portal, while Dark Energy emerges as a result of the spontaneously broken mirror symmetry of the QCD ground-state by means of quantum gravity effects. According to this proposal, neutrinos can also live in a new exotic phase, which is non-relativistic. In particular, cold neutrinos can be produced in the early Universe and undergo condensation exactly due to neutrino interactions with the QCD-induced Dark Energy. In other words, non-relativistic neutrinos re-organize in a coherent state composed by Cooper-like pairs with basic properties of Cold Dark Matter. If we introduce the new Peccei-Quinn symmetry beyond the Standard Model, the notorious QCD axion can be identified with a composite state of neutrinos, thus, naturally resolving the strong CP problem without introducing any new fundamental axion field. Among the phenomenological implications of our scenario, we predict that a photon, in presence of a strong magnetic field, would decay into a pair of neutrinos. This is a phenomenon that might, at least at a first sight, seems bizarre. Nonetheless, within a merely scientific perspective, we are only left with the possibility to check whether this, although being an unexpected but necessary consequence of our unified approach, could actually be realized in Nature. 

\section*{Acknowledgments}
Work of A.A.~is supported by the Talent Scientific Research Program of College of Physics, Sichuan University, Grant No.1082204112427
\& the Fostering Program in Disciplines Possessing Novel Features for Natural Science of Sichuan University,  Grant No. 2020SCUNL209
\& 1000 Talent program of Sichuan province 2021. R.P.~is supported in part by the Swedish Research Council grant, contract number 2016-05996, 
as well as by the European Research Council (ERC) under the European Union's Horizon 2020 research and innovation programme (grant agreement No. 668679).

\bibliographystyle{JHEP}
\bibliography{bib}

\end{document}